\documentstyle[epsf,eqsecnum,aps,prb]{revtex}

\begin{document}
%\draft
%\preprint{PUC-Rio, UFRJ}

\title{Magneto-Roton Modes of the Ultra Quantum Crystal: Numerical Study}

\author{Pascal Lederer $^{+,\diamond}$ \footnote{On leave from 
Physique des Solides,U. P. S., F91405 Orsay, France( Laboratoire
associ\'e au CNRS)} and $^{+}$C. M. Chaves 
 } 
\address{ $^+$Departamento de F\'{\i}sica, PUC-Rio, C. P. 38071, Rio de
Janeiro and $^{+, \diamond}$Instituto de F\'{\i}sica, Universidade Federal
do Rio de Janeiro, C.P.68528, 21945-970, Rio de Janeiro, Brasil} 
\date{\today}
\maketitle
\begin{abstract}
 The Field Induced Spin Density Wave phases observed in quasi-one-dimensional conductors of the Bechgaard salts 
  family under magnetic field exhibit both Spin Density Wave order and a Quantized Hall Effect, which 
   may exhibit  sign reversals. The original nature of the condensed phases is evidenced by the  collective 
   mode spectrum. Besides the Goldstone modes, a quasi periodic structure of Magneto-Roton modes, predicted 
   to exist for a monotonic sequence of Hall Quantum numbers, is confirmed, and a second mode is shown to
   exist within the single particle gap. 
   We present numerical estimates of the 
   Magneto-Roton mode energies in a generic case of the monotonic sequence. The mass anisotropy of the collective 
   mode is calculated. We show how differently the MR spectrum evolves with magnetic field at low and high fields.
      The collective mode spectrum should have specific features, in the sign reversed 
   "Ribault Phase", as compared to modes of the majority sign phases. 
       We investigate numerically  the collective mode
        in the Ribault Phase, and find
 in this latter case a previously unnoticed field independent low energy Magneto-Roton mode, at an intermediate 
wave vector. The occurrence of the Ribault Phase depends sensitively on the electron-electron interactions. A by
 product of our study deals with the metal-Ultra Quantum Crystal instability line: we find, in a monotonic
  sequence, a re-entrant behaviour of the metallic phase, at a given temperature,
as a function of field, which has been observed experimentally. This behaviour also is sensitive to 
the strength of electron-electron interactions. 
  \end{abstract}
\pacs{Pacs numbers 72.15.Nj  73.40.Hm  75.30.Fv. 75.40.Gb}

\widetext
\section{Introduction}
When a static uniform magnetic field is applied perpendicular to their most conducting plane, certain Bechgaard salts
 undergo, at low temperatures, a transition from a metallic state to a cascade of Field Induced Spin Density 
 Wave (FISDW) semi-conducting states; the latter exhibit a Quantized Hall Effect\cite{review,gorkov,gm91,pl96,hlm1}.

 It is not fully recognized yet that this electron hole condensate is an original state of matter, which is neither 
 a conventional Density Wave condensate nor a conventional Quantum Hall electronic liquid. The widely used 
 phenomenological name of the phenomenon ("Field Induced  Spin Density Waves") may be in part responsible for 
 this oversight. This led one  to propose the more correct name of
  "Ultra Quantum Crystal" (hereafter UQC)\cite{pl87}. 
 Indeed, in that system  long range magnetic order (quantum crystal) sets in only because the unpaired carriers may form 
 a few Landau bands
(ultra quantum limit), while those Landau bands
may organize and stay completely filled as the field varies only because magnetic order sets in. 
The Integer Quantum Hall Effect observed in the Bechgaard salts arises {\it because of} electron-electron 
interactions, in a three-dimensional material. This is in sharp contrast with the Integer Quantum Hall Effect 
in the two-dimensional electron gas\cite{prange}.

A clear sign of the unconventional nature of the electron-hole condensate in the UQC is that its condensation
 energy per particle is much smaller than the orbital energy. This precludes a perturbation treatment of the 
 orbital magnetism in the theoretical description of the condensed phase.

Collective modes of the UQC are the clearest evidence for the
specificity of the 
 electron-hole condensate\cite{pl87,dppl,pl97}. Their spectrum exhibits an original structure
in reciprocal space, with a series of local minima at multiples of $G$, the  
magnetic wavevector, 
which recall the local minimum of the Magneto-Roton  (hereafter MR) mode at the inverse magnetic length in the 
Fractional Quantum Hall Effect.\cite{prange}. 
The MR modes  result both from 
the orbital motion of electrons, and from the condensation of a Density Wave order parameter. One might na\"ively think
of them as some sort of Kohn anomaly. However, Kohn anomalies, in quasi 1D conductors,
are phonon local minima in the normal phase, at the wave vector $2k_F$. They become soft at the metal-CDW transition.
 The 
Magneto-Roton modes of the UQC never become soft. They appear at quantized values of the wave vector parallel
component $q_{\vert \vert}=G$, $2G$, or, 
as we shall see for the UQC phases
 called Ribault Phases, and which are characterized by a negative Quantum Hall number, at $nG$,
with $n=4$ to $11$. $G$   is  from three to two orders of magnitude smaller than $k_F$, 
depending on the magnetic field intensity: it does not relate to the Fermi 
Surface, but to the Quantum Hall Effect. It is equal to the change of wave vector from sub-phase to sub-phase and
 not to the subphase Density Wave 
wavevector.

It is a historical fact that, probably because of the advent of high temperature superconductivity, the experimental
 and theoretical effort about the UQC has slowed down considerably since 1986. In particular no experimental study 
 has been published to test the theory of MR modes in the UQC, most of which dates back to 1987. Practically no 
 theoretical progress either has been recorded since the first papers in 1987\cite{pl87} and 1988\cite{dppl}.

Another plausible reason for the lack of experimental effort about the UQC collective modes is the lack of theoretical 
understanding, until recently, of the Quantum Hall sign inversion phenomenon, the so called Ribault anomaly. 
The latter was discovered early on\cite{rib2,pivetau,cooper}, and
was thought by some to be incompatible with the theory of the
"usual" monotonic Quantum Hall Effect and   UQC cascade phase diagram.

As will be discussed further on, the Ribault anomaly is now believed to be 
understandable within the same theoretical scheme, the "Quantum Nesting Model"(QNM), that allows to understand a
good deal of the UQC phenomenology\cite{zm}. As a result, the other theoretical predictions of
the QNM  should be taken seriously and allowed to be tested experimentally in
a detailed fashion.

This paper aims at bridging a gap between the theoretical state of affairs on
the UQC collective modes,  and the need for detailed quantitative predictions which could, and should, be compared
 to experiments. For instance, estimates of MR energies relative to the single particle gap,  predictions about 
 evolution of collective mode energies with magnetic field, estimates of MR effective mass anisotropy,
were non existent until now.

We consider
also  phases  with 
{\it negative} quantum Hall numbers, called "Ribault Phases", which
have a qualitatively different spectrum from sub-phases with {\it positive} (majority sign) QHE\cite{pl97}. 
 In the negative sign QHE phase, MR have  different (larger) {\it wave vectors},  different
 (larger) {\it energies}, and different {\it field dependences} when compared to the MR modes of the majority sign phases. Phases neighbouring the Ribault Phases, either in parameter space or in energy were predicted to have  MR spectra intermediate between those of the normal phases and those of Ribault Phases.

  For the first time detailed numerical estimates of the MR modes (within the
  Random Phase Approximation (RPA)) in the Ribault Phases, and the neighbouring phases  are given, based on model 
  parameter values which describe well the Bechgaard salts. In addition, with the particular set of parameters studied
   here, a new feature, previously unnoticed,
emerges: a third MR mode, with lower energy (in most parts of the phase diagram)
and almost no field dependence is predicted and computed. Furthermore, this study also reveals a large
sensitivity of the sign reversal phenomenon to the strength of electron-electron interactions.

In order to compute the collective mode energies of the ordered phases, a first step deals
with the computation of the metal-UQC instability line. This has yielded an interesting by-product:
 we have found that the QNM describes, on some occasions, a rentrance of the metallic phase at fixed temperature, as 
 a function of magnetic field $H$, which is a well established experimental fact\cite{pesty87}. This rentrance was 
 thought until now to be out of reach of the QNM.

Our paper is organised as follows:  section II summarizes known results on the physics of Bechgaard salts under
 magnetic field. In section III we describe the results about the 
 metal-UQC instability line; we emphasize the sensitivity of the phase diagram to the electron-electron 
 interaction strength, in particular as regards the sign inversion phenomenon. We also discuss
  the re-entrant behaviour mentionned above. A summary of FISDW collective mode theory is given in
  section IV. New results on collective modes are described in section V
  which is divided in two parts: one   describes the  MR spectrum 
  for the usual monotonic sequence of Quantum Hall phases, and
shows how, as the field intensity is stepped up, UQC phases lose their field independent MR modes, and go over at high
 fields to a low energy excitation spectrum
similar to that of a perfectly nested Spin Density Wave. The second part of section V
 gives the corresponding results for "Ribault Phases" 
(phases with Quantum Hall Effect
sign inversion). We show there that the spectrum of MR modes 
exhibits qualitatively different features as compared to the 
monotonic sequence case. The same subsection also discusses the 
features of the MR spectrum in phases neighbouring the 
Ribault Phase in parameter space. 
A summary of the  results and a discussion of their meaning is given in the Conclusion.
 
\section{Summary of Ultra Quantum Crystal physics}

Here we recall in more details some aspects of the physics of Bechgaard salts, and of
their behaviour under magnetic field.

Organic conductors of the Bechgaard salts family, $(TMTSF)_2 X$ where $TMTSF$ =
tetramethylselenafulvalene are  quasi-one-dimensional (quasi-1D) systems. The typical hierarchy of their transfer
 integrals is: $t_a=3000K$,  $t_b=300K$, $t_c=10K$.
In three members of this family ($X=ClO_4,PF_6,ReO_4$), the metallic phase is destroyed by a moderate magnetic field
 $H$ applied along the $c$ direction, perpendicular to the most conducting planes ($a,b$). A cascade of  magnetic 
 phases, separated by first order transitions appears as the field intensity is stepped up: within each sub-phase, 
  a  FISDW 
is stabilized  with a peculiar electronic structure, characterized by a small number of exactly filled Landau levels 
(bands in fact) \cite{pl96}. The Landau bands are separated  by a hierarchy  of gaps $\delta_n$ which oscillate with
 the magnetic field. The phase labelled by $N$ is characterized by 
 the largest gap $\delta_N$ at the the Fermi level. A sum rule $\Sigma_n \delta_n^2= \Delta^2$,
 connects all the gaps to the order parameter
$\Delta$. 

  Each UQC sub-phase exhibits a Quantized Hall conductivity, which is the first example of a Quantum Hall Effect in
   a 3D system.
The cascade of quantized phases results from an interplay between the nesting properties of the Fermi Surface (FS),
 and the quantization of 
electronic orbits in the field: the wave vector of the SDW varies with field so that unpaired carriers in a subphase
 are always organized in
 completed filled Landau bands. As a result the number of carriers in each subphase is quantized, and so is the Hall
  conductivity: $\sigma_{xy}=2Ne^2/\hbar$ \cite{review,pl96}. 

The condensation of the UQC phases results from the peculiar electronic structure of open Fermi Surface metal under
 magnetic field: because of Lorentz force, the electronic motion becomes periodic  and confined along the high 
 conductivity direction of the chains ({\bf a } direction)\cite{gorkov}.

 The periodic orbital motion of the electrons in real space
 is characterized by a wave vector $G=eHb/\hbar$, $b$ being the interchain distance. The orbital wavelength
 $x_0$ is such that the flux threading  the space between two organic chains over a length $x_0$ 
 is one flux quantum $\phi_0$. 
 As a result, the static 
bare susceptibility of the normal 
phase, $\chi _0({\bf Q})$ can be expressed 
as a sum over weighted
 strictly 1D bare susceptibilities which diverge at quantized
  values of the longitudinal component of the wave vector $Q_{n,\vert  \vert}=2k_F+nG$\cite{gm91,pl96,hlm1}.
 The largest divergence signals the appearance of a SDW phase with quantized    
 vector $Q_{N,\vert  \vert}=2k_F+NG$. 

It is important to realize that the periodic electronic motion along the $a$ direction is manifest in a particular
 choice of gauge, the Landau gauge, when
${\bf A_1} =(0,Hx,0)$. With this choice, under the Peierls-Onsager substitution, the transverse electronic dispersion 
relation (which is periodic in $k_yb$), $\epsilon_{\perp}(k_yb,k_zc)$, becomes 
$\epsilon_{\perp}(k_yb-eHbx/\hbar,k_zc)=\epsilon_{\perp}(k_yb-Gx,k_zc)$, i. e. 
 a periodic one-electron potential along $a$. With a different choice of gauge, say ${\bf A_2}=(-Hy,0,0)$,  
 the single electron Schr\"odinger equation is completely different, and no periodic potential appears along $a$, 
 although the 
electronic spectrum is unchanged. This gauge dependent formulation is familiar for problems of orbital magnetism. 
In any particular choice of gauge, the potential vector breaks translational symmetry in a particular way, and 
single electron wave functions are gauge dependent.  Energy spectra and macroscopic response functions are gauge 
independent.

 The  Quantized Nesting Model (QNM)  \cite{hlm1} describes most of the features of the phase diagram in a 
 magnetic field. It
 has been shown recently to explain  the experimental
 observation of  the Hall plateaux sign reversal when the field varies\cite{zm}.
 Most plateaux exhibit the same sign. (By convention we will refer to these plateaux
 as positive ones). The sign reversal has been discovered by Ribault in $(TMTSF)_2ClO_4$
 under certain conditions of cooling rate\cite{rib2}. Negative plateaux
 have been reproduced and also found in $(TMTSF)_2PF_6$ where their existence
 depends crucially on pressure \cite{pivetau,cooper,bali}. Recently,
 Balicas et al \cite{bali} have shown that there exists a range of pressure 
 for which, in the $PF_6$ salt, the sequence of observed plateaux when the field
 decreases can be identified with the quantum numbers $N=1,2,-2,3,4,5,6,7$.
 A more ancient experiment has shown a sequence of phases $N=1,2,-2,4,-4,5,6$\cite{cooper}. Hereafter, we  will
  refer to the UQC Phase with negative Hall number as "Ribault Phases".

 Zanchi and Montambaux\cite{zm} have shown that the negative plateaux can be understood within the QNM assuming 
 the dispersion relation in the normal phase to be:

\begin{eqnarray} \label{model}
\epsilon({\bf k})& =& v_F(\vert k_x\vert - k_F)+ \epsilon_{\perp }({\bf k_{\perp}}),\\
\epsilon_{\perp }({\bf k_{\perp}})& =& -2t_b\cos k_yb  -2t_c\cos k_zc -
2 t'_b\cos 2k_yb \nonumber \\ &&-2t_3\cos3k_yb -2t_4\cos4k_yb
 \nonumber
\end{eqnarray}

$\epsilon(k_{\perp})$is a  periodic function which describes a warped FS.
With $t_3=t_4=0$,  Eq.(\ref{model} ) cannot lead to sign reversals, as 
$sign(N)=sign(Q_{\vert  \vert}-2k_F)=sign(t'_b)$ \cite{zm}.  Small values of $t_3\simeq .2t'_b=2$K, and $t_4=.2$K,
 however, are sufficient to account for the experimental results of Balicas et al. 
 (A slightly different explanation for the negative Hall Effect phases  has been put forward recently, 
 based on the consideration of Umklapp terms \cite{yakonico} and a non-zero $t_4 $ term. 
 We will not consider the Umklapp terms in this work.)

In Eq.(\ref{model}), the only parameters which are not determined yet by experiments are $t_3$ and $t_4$. Those parameters certainly exist as a correction to the linearized form of the dispersion relation along $a$, and they
have to be small compared to $t'_b\simeq 10K$.

The  normal metal-FISDW instability line  $T_{cN}(H)$ is given by: 
\begin{equation} \label{ki}
\chi _0({\bf Q},T_{cN}, H)=
\Sigma _nI_n^2 (Q_{\perp })\chi _0^{1D}(Q_{\vert  \vert}-2k_F-nG, T_{cN})=1/\lambda
\end{equation}
$\lambda$ is the electronic interaction constant. Eq.(\ref{ki})
 exhibits the structure of $\chi _0$ as the sum of one
 dimensional terms $\chi _0^{1D}$ shifted by the magnetic field wave
 vector $G=eHb/\hbar$. 
$\chi_0^{1D}\propto -\ln(\max \{ v_F(2k_F-q),T   \}/\epsilon_F)$. 
In Eq. (\ref{ki}),  the coefficient $I_n$ depends on the dispersion  relation and H:
\begin{eqnarray}
I_n(Q_{\perp})  & =& 
<\exp i\left[ T_{\perp}(p+Q_{\perp}/2)+ 
  T_{\perp}(p-Q_{\perp}/2))+np
 \right] > 
\end{eqnarray}
where $T_{\perp}(p) = (1/\hbar \omega _c)\int_0^p\epsilon_{\perp}( p')dp'$
and $<...>$ denotes the average over p.

\section{The metal-UQC instability line}

\subsection{The sequence of UQC phases }
Eq.(\ref{ki}) allows to compute the metal-UQC instability line. We show here that it is not sufficient,
 to determine the sequence of UQC phases, to examine 
the maxima of $\chi_0(\bf q)$ as a function of ${\bf q}$ at different fixed temperatures .
This remark is especially meaningful in the description of the sign inversion
phenomenon: depending on the value of $\lambda $ in (\ref{ki}) different sequences of UQC phases emerge. 

Define $T_{\infty}$ as the FISDW ordering temperature in infinite field. In the latter 
case $I^2_n=\delta_{n,0}$ and $\chi(q_{||}=2k_F, q_{\perp=\pi/b})\propto \ln{2\gamma E_0/\pi T}$. From Eq.(\ref{ki}) ,
we have $T_{\infty}=(2\gamma/\pi)E_0 \exp(-1/\lambda)$. ($\gamma$ is Euler's constant and $E_0$ a high energy cut-off). $T_{\infty}$ is equal to the ordering temperature for the perfectly nested Fermi Surface at any field (i. e. when $t'_b=0$). For a given value of the cut-off parameter $E_0$, changing $\lambda$ is equivalent to changing $T_{\infty}$.
 It is clear from Eq.(\ref{ki}) that the instability line
 results from the largest solution $T_{cN}$ at a given field, among all solutions 
 obtained by varying $\bf q$.  Now define a generalized instability temperature
  $T_{N \pm m}(\pm q_{\perp})$ as a lower temperature solution,  i. e. one that corresponds 
  to a larger free energy than the physical one; it corresponds, at a given field, to a different 
  wave vector from that of the 
actual instability; for any integer m such that $q_{||}-2k_F= (N+m)G$, a whole family of
 solutions is obtained by varying $q_{\perp}$:

\begin{equation} \label{Tc}
   \frac{ T_{N \pm m}}{T_{\infty}}     = \exp \left[                  
\sum_{n \neq  0} \frac{ I^2_{N\pm m +n} (Q_{\perp }^N \pm q_{\perp})}{  I^2_{N\pm m}(Q^N_{\perp }
 \pm q_{\perp }) }\ln \left(    \frac{\pi T_{\infty}}{2\gamma \vert n\vert \omega _c}\right) \right]
\end{equation}
In Eq.(\ref{Tc}), we have used the sum rule: $\Sigma_n I_n^2=1$. Eq.(\ref{Tc}) is
valid for all $k_BT<\hbar \omega_c$.
For $m=0$ and $q_{\perp}=0$, $T_{N \pm m}=T_{cN}$, the ordering
 temperature for the $N$th subphase. For $m\neq 0$, $T_{N \pm m}(q_{\perp})$ 
 generalizes the definition of the critical temperatures on either side
 of phase $N$ in the $(T,H)$ plane. $T_{N \pm m}(q_{\perp})$ are at
 most equal to the virtual transition lines $T_{ N\pm m}$ which can 
be drawn in the $N$th subphase part of the phase diagram\cite{pl87}. In the ($T,H$)
 plane, there is an infinite number of continuous lines  crossing 
the phase diagram. The upper limit of this family is the actual
 (continuous non analytic) transition line from the normal metal
 to the UQC; this  line coincides piecewise with the transition lines
 labelled by  the successive integers describing the Quantum Hall
conductivity\cite{lm}.

It is clear from inspection of Eq.(\ref{Tc}) that {\it $ T_{cN}$ does not scale with $T_{\infty}$},
 since the ratio $T_{cN}/T_{\infty}$ on the left hand side of
Eq.(\ref{Tc}) depends on $T_{\infty}$ through the argument of the logarithms on the right hand side.
 The variation of the ratio $T_{cN}/T_{\infty}$ with $T_{\infty}$ is
complicated, with no clear tendency emerging at first sight because the coefficients $I_{N \pm m}$
 have an oscillating behaviour as a function of their arguments $t_i/\hbar \omega_c$ .  
 Numerical results exhibit a rather sensitive dependence of
the Quantum Hall sequence and of its sign changes with $T_{\infty}$, which is one of the new results of this paper.

 $T_{\infty}$ is not a free parameter: in our case, we are interested in a metallic ground state
 in the absence of field, therefore it cannot exceed $t'_b/k_B$, where $t'_b$
 violates the perfect nesting condition. $t'_b$ has to exceed the value of the zero field SDW ordering temperature 
 of the perfect nesting case, which is precisely equal to $T_{\infty}$.

\begin{figure}[1a]
 \begin{center}
\epsfysize=8cm
\epsfxsize=8cm
\epsfbox{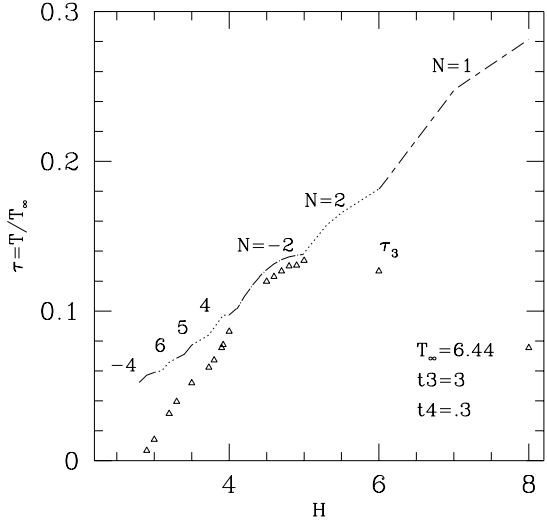}
\end{center}
\caption{Metal-FISDW instability line $T_{cN}$ as a function of field for $T_{\infty}=6.44$K.
 The values
 of the small parameters $t_3=3$K and $t_4=.3$K are the same in Fig.(\ref{Fig:1a}) 
 and Fig.(\ref{Fig:1b}). Notice the absence  of phase $N=3$.
 }
 \label{Fig:1a}
            \end{figure}

\begin{figure}[1b]
 \begin{center}
\epsfysize=8.6cm
\epsfxsize=8cm
\epsfbox{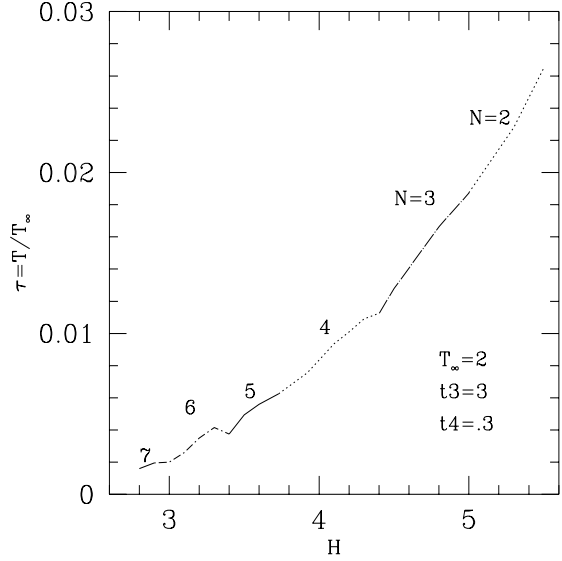}
\end{center}
\caption{Metal-FISDW instability line $T_{cN}$ as a function of field for $T_{\infty}=2$K.  Substantial changes in the 
FISDW sequence of phases are observed in comparison with Fig.(\ref{Fig:1a}). The sequence of phases
is monotonic (positive). The values
 of the small parameters $t_3=3$K and $t_4=.3$K are the same in Fig.(\ref{Fig:1a}) and Fig.(\ref{Fig:1b})
 }
 \label{Fig:1b}
            \end{figure}

 The results of the numerical analysis for the transition line for different values of $T_{\infty }$ 
 are shown on Fig.(\ref{Fig:1a}) and Fig.(\ref{Fig:1b}) for two different choices of the parameters $t_3$ and $t_4$.
       \   
       \   
  The dependence of the phase diagram is particularly clear when $t_3, t_4$ are both small compared to $t'_b$,
as they should. In this latter case, the sign inversion phenomenon disappears altogether when $T_{\infty}$
 is changed by a factor 3, from $T_{\infty}=6.44$K to
$T_{\infty}=2$K. The former value was used in previous work\cite{dp87} (see Fig.(1)). There is no sign inversion
 phenomenon when $T_{\infty}=2$ K, while the sequence of phases for $T_{\infty}=6.44$ K is : $-4,6,5,4,-2,2,1$. 
 Note that phase $-4$ becomes stable, when the field increases, {\it before} phases $6$ and $5$, an
unusual inversion of Hall number absolute magnitude which has not yet been observed experimentally. Note also that the
 decrease of $T_{\infty}$ by a factor 3
triggers a decrease of ordering temperatures by  roughly two orders of magnitude between $H=2.8 $ T and $H=5.3$ T.

\begin{figure}[2a]
 \begin{center}
\epsfxsize=8.6cm
\epsfysize=8cm
\epsfbox{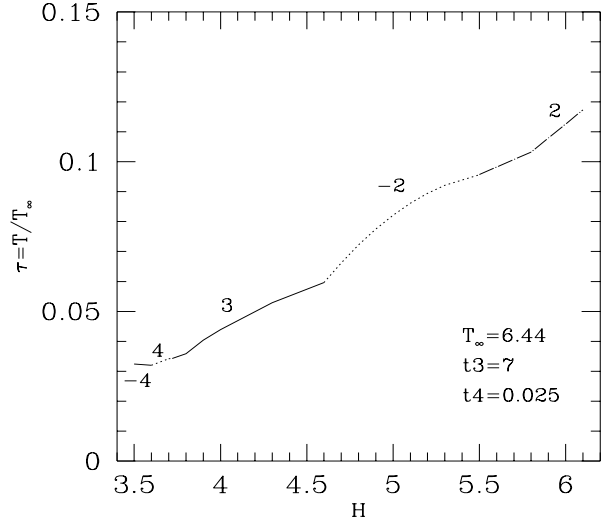}
\end{center}
\caption{Metal-FISDW instability line $T_{cN}$ as a function of field for $T_{\infty}=6.44$K.  Substantial changes in the 
FISDW sequence of phases are observed in comparison with Fig.(\ref{Fig:1a}). The values
 of the  parameters $t_3=7$K and $t_4=.025$K are the same in Fig.(\ref{Fig:2a}) and Fig.(\ref{Fig:2b})
 }
 \label{Fig:2a}
            \end{figure}

\begin{figure}[2b]
 \begin{center}
 \epsfxsize=8.6cm
 \epsfysize=8cm
 \epsfbox{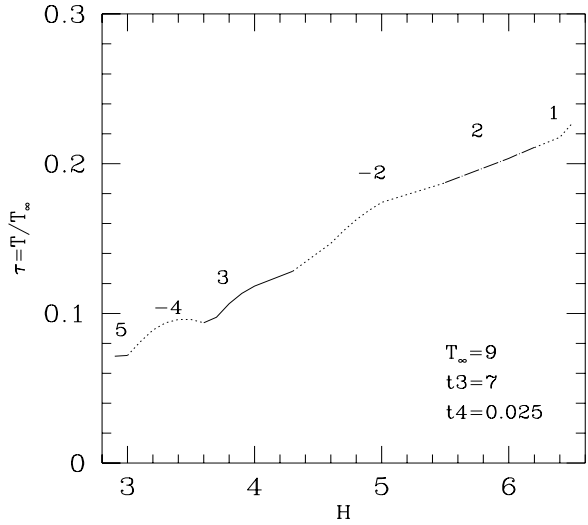}
\end{center}
\caption{Metal-FISDW instability line $T_{cN}$ as a function of field for $T_{\infty}=9$K.  
 In comparison with Fig.(\ref{Fig:1b}), phase $N=4$ has vanished from the phase diagram. The values
 of the  parameters $t_3=7$K and $t_4=.025$K are the same in Fig.(\ref{Fig:2a}) and Fig.(\ref{Fig:2b})
 }
 \label{Fig:2b}
            \end{figure}

For the choice of values $t_3=7$ K and $t_4=.025 $ K (taken from the literature\cite{zm}), the phase diagram also
 exhibits large changes upon a change of $T_{\infty}$ from $6.44 $ K to $9$ K.
  See Fig.(\ref{Fig:2a}) and Fig.(\ref{Fig:2b}). The sequence of phases 
 $5,-4,3,-2,2,1$ obtained at $T_{\infty}=9$ K  changes to the following: $5,-4,4,3,-2,2,1$ when 
 $T_{\infty}\rightarrow 6.44$ K: all other things equal, the change of $T_{\infty}$ by less than 
 $30 \%$ triggers the disappearance of phase $4$, as well as changes of ordering temperatures by $600 \%$ to $200 \%$.
      \   
       \   
 The sensitivity of the phase diagram, and especially of the magnitude
 of the ordering temperatures on $T_{\infty}$ is such  that knowing  the phase Quantum Hall
 number and the value of the transition temperature  at a given pressure
 allows a rather reliable experimental value for $T_{\infty}$.
 
\subsection{Re-entrance of the metallic phase }

\begin{figure}[3a]
 \begin{center}
 \epsfxsize=8.6cm
 \epsfysize=7cm
 \epsfbox{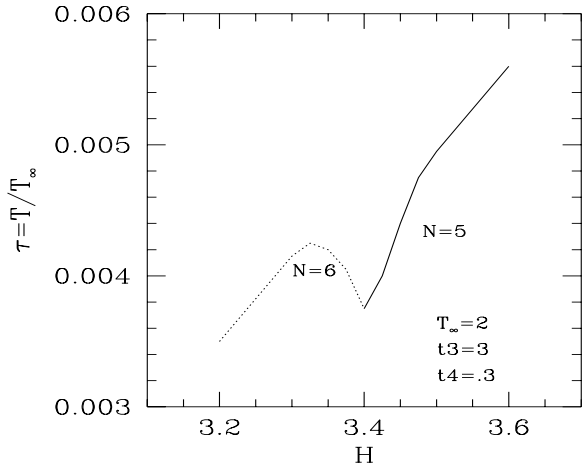}
 \end{center}
 \caption{Re-entrance of the Metal-FISDW instability line $T_{cN}$ as a function of field for $T_{\infty}=2$K.  
 Here $t_3=3$K and $t_4 =.3$K. However the re-entrant behaviour also exists within the model with zero
 value for both those parameters. See Fig.(\ref{Fig:3b})
 }
 \label{Fig:3a}
            \end{figure}

\begin{figure}[3b]
\begin{center}
\epsfxsize=8.6cm
\epsfysize=7cm
\epsfbox{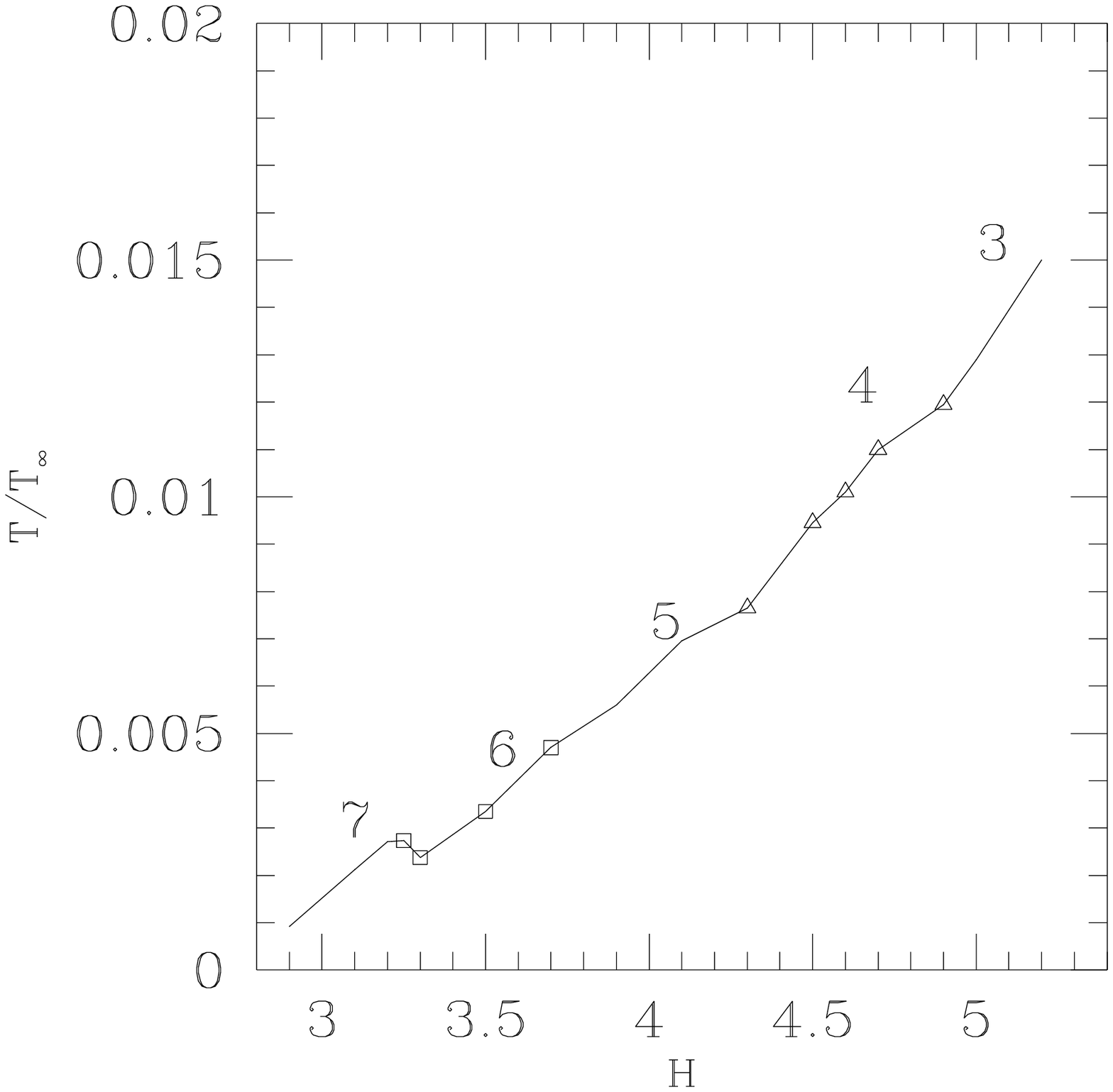}
\end{center}
\caption{Re-entrance of the Metal-FISDW instability line for $T_\infty=2$K
and $t_3=t_4=0$K
 }
 \label{Fig:3b}
            \end{figure}

Detailed experimental determination of the instability line has revealed that  $ dT_c(H)/dH$ may change sign as 
 a function of $H$\cite{pesty87}.
 This means that
the metallic phase, at a fixed temperature exhibits a re-entrant behaviour as a function of field.
  This feature was thought to be outside the reach of the QNM in its usual formulation
 for the monotonic Quantum Hall number sequence because, until now
 the theoretical
 derivations of the transition lines
within the QNM with $t_3=t_4=0$  have not reproduced this feature.
  We have 
first found this behaviour in the numerical simulations with non 
zero values for $t_3$ and $t_4$, as shown in Fig.(\ref{Fig:3a}).
We have then studied this phenomenon as a function of $t_3$ and $t_4$. The outcome of 
this study is that re-entrance
appears already within the model with $t_3=t_4=0$, within the framework of 
the monotonic sequence of Hall numbers,
 as shown on Fig.(\ref{Fig:3a}) and Fig.(\ref{Fig:3b}).
It appears to depend also on the value of $T_{\infty}$. 

Our conclusion about this aspect of the phase diagram is that there 
is nothing mysterious about
the re-entrant behaviour, which is well within reach of the simplest
 QNM; it is a simple consequence of the
general shape of the $T_N(H)$ curves as function of field: each in turn
 goes through a maximum then decreases with field. If for a 
 given $N$ the maximum as a function of field is also the absolute maximum for
 all $N$ values and for the same field, re-entrant behaviour follows.
 
\section{Summary of FISDW Collective Mode Theory}
We now turn to  the collective modes of the UQC
which depend directly on the
 network of real and virtual transition lines defined in the last section. Using this network for the monotonic 
 Quantum Hall sequence, ($t_3=t_4=0$), Lederer and Poilblanc
showed  that  the UQC collective modes exhibit, aside from the usual Goldstone modes linear in wave vector, at 
least one Magneto-Roton (hereafter MR) mode within the single particle gap, located at
 $q_{\vert  \vert}=G, 2G,...$, $q_z=\pi /c$, and  some optimum $q_y$\cite{pl87,dppl}. Only the 
  mode at $q_{||}=G$ was actually proved to exist. Together with the usual Goldstone modes, it is the signature
   of the novel nature of the electron-hole condensate in UQC: a Spin Density Wave driven by orbital quantization 
   and an Integer Quantum Hall system driven by electronic interactions. The MR mode is both a consequence of 
    quasi 1D electronic periodic orbital motion, gauge invariance, and the long range crystalline order of the 
    Field Induced Phase\cite{dppl}.

The physical mechanism behind the MR minimum may be described as follows: the Goldstone boson at wave vector
 $q_{||}=mG$, in a usual SDW phase, is an overdamped mode, because $mv_FG
=2m\hbar \omega_c$, which is much larger than the single particle gap, at any value $m\geq 1$\cite{note}. However,
 a fluctuation mode, in phase $N$, at a wave vector 
$mG$ can scatter against the self consistent periodic orbital potential (responsible for the gaps at $\pm (k_F +mG)$
 in the single particle dispersion relation) and shift its parallel momentum by $|mG|$, so as to propagate as a virtual 
 mode with $q_{||}=0$, i. e. a zero energy mode! However, it is not really a zero energy mode, as it would if
  the periodic orbital potential was a bona fide electrostatic potential, with bona-fide Bragg reflections. In 
  fact this virtual shift in wave vector allows the mode
at $q_{||}=mG$ to propagate in a medium which is not exactly like phase $N$ for $q_{||}=0$, but which is a mixture 
of unstable (excited) phases $N\pm m$. As we shall see, the closer the free energies of phases $N\pm m$ to that of
 phase $N$, the lower the MR energy within the single particle gap.

On the other hand, once the  Quantum Hall sign inversion phenomenon is explained within the QNM, at the cost of 
refining the description of the electronic dispersion relation, the  collective mode theory derived for the monotonic
 UQC sequence applies directly to the "Ribault Phase", the phase with inverted Quantum Hall sign\cite{pl97}. This 
 has a surprising consequence:  exploiting the topology of the phase diagram in the presence of Hall Effect sign changes, 
 one shows that the MR spectrum for the Ribault Phase $-2$
  should be qualitatively different from the usual  spectrum in the monotonic sequence case, so that the Ribault Phases
   are physically distinct from the majority sign UQC phases, and not just the same objects 
   with negative  carriers sign. For the particular sequence observed by Balicas
    et al.\cite{bali}, i.e.  1, 2, -2, 3, 4, 5, 6, 7 , the differences in the -2 phase as opposed 
    to the usual UQC subphase are the following:
\begin{itemize}
\item there are no MR modes with energy minimum at $q_{||}=G$ or at $q_{||}=2G$
\item  there are at  two MR modes with energy minimum at $q_{||}=4G$ and $q_{||}=5G$
\item These MR modes have substantially higher energies than their usual counterparts. In particular, within 
the RPA, their energies are larger than half
the single particle gap
\item Both energy minima have a much larger field dependence than their normal
counterparts
\item those field dependences have opposite signs: the mode at $q_{||}=4G$ is a
decreasing function of the magnetic field, while the mode 
at $q_{||}=5G$ is an increasing function of the magnetic
 field 
\end{itemize}  

Likewise, the MR spectrum of UQC phase $N$ ($N>0$)
 neighbouring  the Ribault Phase $- |P|$  are
 "contaminated" by their neighbour, with one usual MR mode at $q_{||}=G$ 
 and one  mode at larger wave vector
  $q_{||}=mG=(|P|+N)G$ , with large magnetic field dependence.
    This behaviour will be understood qualitatively from the equation for the mode energies in 
    the following subsections;
  as for the modes within the Ribault phase, it is due to the 
  dissymetry between virtual phase energies
  for phases $N\pm m= N\pm (N+ |P|)$ since $N+m$ and $N-m$ have opposite signs. 
  Modes at $G$ and $2G$ for $N>1$  mix virtual phases $N\pm 1$ and $N\pm 2$ which both have
  free energies nearly equal to that of phase $N$. In contrast,
   for a phase with $N<0$, the mode at $mG$ mixes
  a low free energy virtual phase $N+m$ ($N+m>0$) with a large free energy one $N-m$ ($N-m <0$).

\subsection{The dynamic spin-spin correlation function}

Technically MR energies have  been derived within the RPA\cite{pl87,dppl},
 by looking at the poles of the spin-spin correlation function
 of the ordered phase. 

This can be shown by explicit microscopic computation of the
transverse dynamic spin-spin correlation function $\chi_{+-}(q,\omega)$ in
the RPA, by extending the work of Lee Rice and Anderson on collective
modes in CDW phases\cite{lra}. 

Letting
$$\tilde\sigma_+=\psi^+_{2\uparrow}(\bar
x)\,\psi_{1\downarrow}(\bar x)$$
and
$$\tilde\sigma_-=\psi^+_{1\downarrow}(\bar
x)\,\psi_{2\uparrow}(\bar x)$$
where $\psi^+_{\alpha\sigma}(\bar x)$ is the creation field operator for a
particle of spin $\sigma$ on sheet $\alpha$ of the Fermi surface, at
space time coordinate $\bar x\equiv(\vec r,t)$; then we can define

$$\chi_{\alpha\beta}=-\langle T_\tau\tilde\sigma_\alpha(\bar
x)\,\tilde\sigma _\beta(\bar x')\rangle\qquad \alpha\ne\beta$$

$$\Gamma_{\alpha\alpha}=-\langle T_\tau\tilde\sigma_\alpha(\bar
x)\,\tilde\sigma _\alpha(\bar x')\rangle e^{i\alpha\vec Q_N(\vec
x+\vec x\,')}$$

The dynamic susceptibility in the ordered phase, as well as the off
diagonal response function $\Gamma_{\alpha\alpha}$ can be obtained
within the RPA, from the non interacting response function, defined
as follows
$$\chi^{0F}_{+-}(\bar x,\bar x')=G_{22}(\bar x,\bar x')\, G_{11} (\bar
x',\bar x)$$
$$\Gamma^{0F}_{++}(\bar x,\bar x')=G_{21}(\bar x,\bar x')\, G_{21} (\bar
x',\bar x)$$
$G_{jj}(\bar x,\bar x')$ is the propagator for the field operator on
side $j$ of the Fermi surface. $G_{ij}(\bar x,\bar x')$ (with $i\ne
j)$ is the anomalous propagation in the presence of the non zero
order parameter. Here 1 and 2 represent $1\uparrow$ and $2\downarrow$.
The RPA yields the following expression
\begin{eqnarray}
&&\chi^F_{+-}(\vec Q=\vec Q_N+\vec q,\omega)=\nonumber\\
&&{\chi^{0F}_{+-}(\vec Q_N+\vec q,\omega)-\lambda
\left[\chi^{0F}_{+-} (\vec Q_N+\vec Q,\omega)\, \chi^{0F}_{+-}(\vec
Q_N,\vec q,\omega)-\Gamma^{0F}_{--}(\vec q,\omega)\,\Gamma^{0F}_{++}(\vec
q, \omega)\right]\over \left[1-\lambda \chi^{0F}_{+-} (\vec Q_N-\vec
q,\omega)\right] \left[1-\lambda \chi^{0F}_{+-} (\vec Q_N+\vec
q,\omega)\right] -\lambda^2\Gamma^{0F}_{--}(\vec
q,\omega)\,\Gamma^{0F}_{++}(\vec q, \omega) }\nonumber\\
&&\nonumber\\
&&\Gamma^F_{--}(\vec q,\omega)=\nonumber\\
&&{\Gamma^{0F}_{--}(\vec q,\omega)\over \left[1-\lambda \chi^{0F}_{+-}
(\vec Q_N-\vec  q,\omega)\right] \left[1-\lambda \chi^{0F}_{+-} (\vec Q_N+\vec
q,\omega)\right] -\lambda^2\Gamma^{0F}_{--}(\vec
q,\omega)\,\Gamma^{0F}_{++}(\vec q, \omega) }\nonumber
\end{eqnarray}
Those formulas are quite general. In the usual case of SDW in zero
magnetic field, they simplify because of the symmetry $\vec q\to
-\vec q$, i.e. $\chi^{0F}_{+-}(\vec Q_N-\vec q,\omega)=
\chi^{0F}_{+-}(\vec Q_N+\vec q,\omega)$, so that
$$\chi^F_{+-}(\vec Q_N+\vec q,\omega)\pm\Gamma^F_{--}(\vec q,\omega)={
\chi^{0F}_{+-}(\vec Q_N+\vec q,\omega)\pm\Gamma^{0F}_{--}(\vec
q,\omega)\over 1-\lambda( \chi^{0F}_{+-} (\vec Q_N+\vec
q,\omega)\pm\Gamma^{0F}_{--}(\vec q,\omega))}$$

The latter expression leads to the phase and amplitude modes of the
SDW, once one uses the gap
equation 
$$\chi^{0F}_{+-}(\vec Q_N,\omega=0)-\Gamma^{0F}_{--} (\vec q=0,
\omega=0) ={1\over \lambda}$$
The simplified expression is still approximately valid, in the limit
$|\vec q|\ll{2\pi\over x_0}$, for the Ultra Quantum crystal. For a
general $\vec q$, however, the correct collective mode equation is
given by the poles of $\chi^F_{+-}(\vec Q_N+\vec q,\omega) \pm
\Gamma^F_{--} (\vec q,\omega)$, i.e. by the equation
$$\left[1-\lambda \chi^{0F}_{+-} (\vec Q_N-\vec q,\omega)\right]
\left[1-\lambda \chi^{0F}_{+-} (\vec Q_N+\vec q,\omega)\right] = 
\lambda^2\Gamma^{0F}_{--}(\vec q,\omega)\,\Gamma^{0F}_{++}(\vec q,
\omega) $$

\subsection{Collective mode equation}

The electronic orbital motion, together with the RPA treatments of interactions,  results in 
effective scattering potential energy terms which
couple electron states with wave vector $k_{\vert \vert}$
 and $k_{\vert \vert}+2k_F +nG$, (n integer) on either side 
of the Fermi Surface. This results in a series of gaps in 
the condensed phase dispersion relation\cite{pl96}, corresponding
 to the various potential scattering terms. The simplest 
approximation which captures the esssential physics resums to 
all orders the gap $\delta_N= I_N \times  \Delta$ at the Fermi level
 and takes all other gaps into account to second order in perturbation\cite{pl87,dppl}. The limit
  $2\delta_n/\omega_c\ll 1$
 is assumed to hold.
   Then the  equation for collective modes 
in the UQC phase $N$ reduces to:

\begin{eqnarray} \label{rot2}
\left( \ln \left(  \frac{2\gamma E_0}{\pi T_{N+m }} \right) -   \tilde{\bar{\chi_0}}(\delta,\omega)                                             \right) 
 \left(  \ln \left(  \frac{2\gamma E_0}{\pi T_{N-m}} \right) -\tilde{\bar{\chi_0}}(\delta, \omega)   \right) &  
=&\left( \tilde{\bar {\Gamma_0}} (\delta ,\omega )    \right)^2 
\end{eqnarray}

where $T_{N \pm m}$ is defined in Eq.( \ref{Tc}), and 
$q_{\vert  \vert}=2k_F +(m +\delta)G$, with $ \delta\ll 1$.

$\tilde{\bar{\chi_0}}$ and $\tilde{\bar{\Gamma_0}}$ are for $n=0$ the
 objects discussed in \cite{lra} in connections with collective modes of SDW. 
Eq.(\ref{rot2}) finally yields, setting $x_{N,m}=\omega(q_{\vert \vert}/G=m+\delta, q_\perp)/(2\delta_N)$,
 (with $q_z=\pi/c$):
\begin{eqnarray} \label{Tm}
\left(\ln \left(\frac{T_{cN}}{T_{N+m}}\right)   -(x_{N,m}^2-1/2)h(x_{N,m}) \right) 
 \left( \ln\left(  \frac{T_{cN}}{T_{N-m}}\right)-(x_{N,m}^2-1/2)h(x_{N,m}) \right) &=&h^2(x_{N,m})/4   
   \end{eqnarray}
with
$h(x,T ) =
\int_0^\infty du\frac{\tanh (\frac{\delta_N}{2T}\cosh u)}{\cosh^2u-x ^2}$. 
In the limit $T=0$K, $h(x,T\rightarrow 0)$
reduces to $$
h(x)=\frac{sin^{-1}x}{x(1-x^2)^{1/2}},  \; \; \; \; x<1
$$
\subsection{Analytical results}
Define $\epsilon_{\pm m}=(T_{cN}-T_{N\pm m})/T_{cN}$

Simple solutions are found for Eq.(\ref{Tm}) in the limit that $\epsilon_{\pm m}\ll 1$. Then
$$x_{N,m}^2\simeq(\epsilon_{-m} + \epsilon_{+m})/2$$.
Since $\epsilon_{-m}$ and $\epsilon_{+m}$ have roughly equal and opposite variations
with magnetic field within phase $N$, the MR minimum is roughly field independent.
In general, in most of the parameter space of a given phase, $\epsilon_{\pm m}>.3$ except
for $m=1$, so that accurate results demand a numerical solution. In particular 
a numerical analysis is required to determine whether a solution
 exists or not within the single particle gap for $m>1$. 
 In order to find the lowest energy solution, one must vary the momentum transverse
 component in Eq.(\ref{Tc}).  However, another simple 
solution holds in the case that $\epsilon_m\ll 1$, while $\epsilon_{-m}\simeq 1$. This situation, as
we shall see, is of interest in the case of the Ribault Phase. It is then straightforward to show from
Eq.(\ref{Tm}) that $x_m^2\simeq 1/2 +(6/\pi)\epsilon_m -\pi/(ln(T_{cN}/T_{-m}))\simeq 1/2 $.
In most other cases, a numerical solution of Eq.(\ref{Tm}) proves necessary.

The following section presents the results of our detailed numerical analysis.

\section{The spectrum of MR modes}
\subsection{ The monotonic Quantum Hall number sequence}
We now exploit the results of the last section. First we derived, at $T=0K$, the MR mode spectrum in a generic case:
 a phase well within a monotonic sequence of UQC phases. We picked phase $N=4$. Then we study the differences which 
 arise at high field, for the last two phases of the monotonic sequence: $N=1$ and $N=0$.
\subsubsection{The generic case}

 The parameters we took for the numerical estimates are relevant to the Bechgaard salts, and have been used
  previously in the literature\cite{lm}: $t_a=3000K$, $t_b=265K$, $t'_b=10K$, $T_{\infty}=6.44$. 
(Those parameters may change with pressure, notably $t'_b$).

\begin{figure}[mr2]
 \begin{center}
 \epsfxsize=8.6cm
 \epsfysize=8cm
\epsfbox{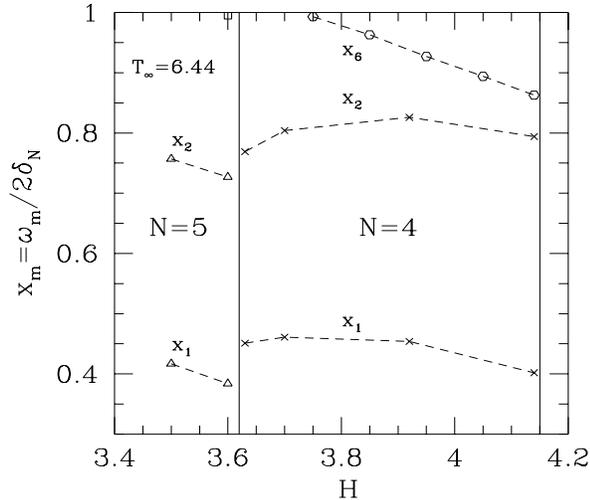}
\end{center}
\caption{ Energy minima of the MR modes in phase $N=4$ of the monotonic Quantum 
Hall sequence ($t_3=t_4=0$K).
The lowest energy mode has $q_{\vert \vert}=G$, the intermediate energy MR mode has 
$q_{\vert \vert}=2G$, the highest energy mode is field dependent and has $q_{\vert \vert}=6G$ 
and couples to the virtual phase at $N=-2$. It is due 
to "contamination" of the Ribault phase at $N=-2$
}
\label{Fig:mr2}
            \end{figure}

 The results are  shown on Fig.(\ref{Fig:mr2}). We find two modes, at $q_{\vert \vert}=G$ 
 or $2G$ which vary little with $H$,
and a {\it third mode}, rather close to the upper edge of the gap, at {\bf $q_{\vert \vert }=6G, q_\perp=0$}. 
This mode is yet another example of "contamination" by the Ribault phase $N=-2$ discussed in the next subsection
 \cite{pl97}. 
It has a monotonic $H$ negative variation, and drowns into the single particle continuum for $H\lesssim 3.75$T.

 We have looked for MR mode with $q_{\vert \vert}=8G$, in phase $4$, which 
 might arise from contamination of the 
 near-by Ribault phase $N=-4$, and found it did not exist in the absence 
 of additional terms in Eq.(\ref{model}),
  but this mode will certainly become alive under pressure\cite{zm,pl97}.

\begin{figure}[z]
   \begin{center}
   \epsfxsize=8.6cm
   \epsfysize=8cm
   \epsfbox{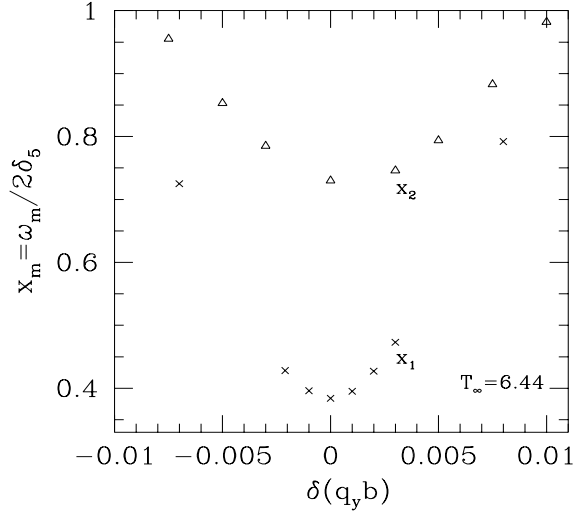}
   \end{center}
   \caption{Dispersion of the MR modes in the transverse direction for the two low 
   energy modes in UQC phase $5$; crosses correspond to the mode with $q_{\vert \vert}=G$;
    triangles 
   correspond to the mode with $q_{\vert \vert}=2G$; 
results are similar for phase $4$. The effective mass $m_\perp$ is found to be about 
$10^3$ larger than $m_{\vert \vert}$. }
\label{Fig:mr3}
            \end{figure}

 The RPA evaluation yields
 a MR minimum $x_{5,1}^{min}\simeq .4$, located at $q_\perp^{min}=.0178/b$
 for $H=3.6T$.
 By varying $q_{\perp}$ around $q_{perp}^{min}$ at fixed $q_{\vert vert}$,
  we can compute the transverse MR mode dispersion.
  We find  a {\it very large} mass anisotropy 
 $m_{\perp}/m_{\vert \vert}|_{N=5,m=1}\simeq .9 \times 10^3$, with
 $m_{\vert \vert}= \hbar \omega_{min}/v_F^2$. The fact that
 $x_{5,1}<1/2$ makes this MR mode a good candidate for the
 kind of specific heat anomalies discussed in ref(\cite{pl96}).
 The mass anisotropy is larger for the mode $q_{\vert \vert}=2G$,
 we find $\simeq$ $2.25\times 10^3$.See Fig.(\ref{Fig:mr3}).  The MR energy $\omega_{Nm}^{min}$
  is discontinuous at the 
 transition from $N$ to
 $N\pm 1$. This contributes a (small) term, which corrects the mean field estimate
  to the transition latent
  heat.  
 On the other hand, $\omega^{min}_{4,6}$ goes continuously at the transition
  to $N=3$ into $\omega^{min}_{3,5}$
  (i. e. a {\it distinct} MR with the {\it same} energy). The computed MR
   dispersion relation is shown on Fig.(\ref{Fig:mr3}).
   
\subsubsection{Large magnetic fields}
\begin{figure}[mr4]
\begin{center}
\epsfxsize=8.6cm
\epsfysize=8cm
\epsfbox{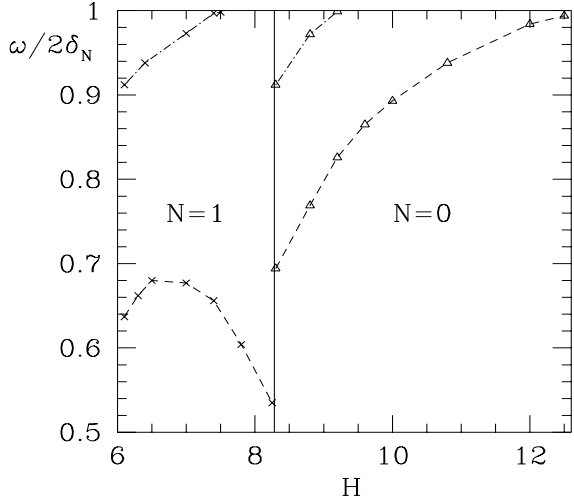}
\end{center}
\caption{Variation of the MR energy minima with magnetic field at large fields. 
The lowest MR mode at $q_{\vert \vert}=G$ in phase $N=1$ has a qualitatively different behaviour
as a function of field than the mode at $2G$ in phase $N=1$ and those at $G$ (low energy) and $2G$
(high energy) in phase $N=0$. This behaviour stems from the network of real and virtual
 transition lines  depicted in Fig.(\ref{Fig:mr1})}
 \label{Fig:mr4}
            \end{figure}

\begin{figure}[mr1]
\begin{center}
\epsfxsize=8.6cm
\epsfysize=8cm
\epsfbox{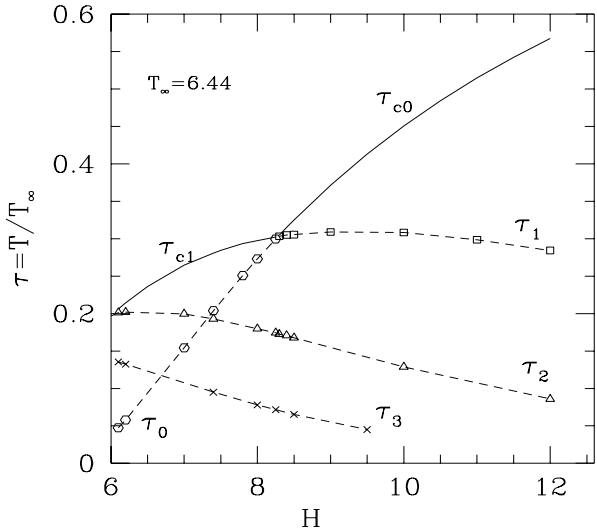}
\end{center}
\caption{Network of real and virtual transition lines of the UQC phase diagram at large fields. For
$H> 8.1$T, the free energy difference between the thermodynamically stable
 phase $N=0$ and the unstable phases 1 and 2 increases rapidly with field. }
 \label{Fig:mr1}
            \end{figure}

The situation changes for $H>6.1$ T, the critical value for the transition 
from UQC phases $2$ to $1$. The reason for this change is apparent from the $T(H)$ curve in Fig.(\ref{Fig:mr1}):  
in phase 1,
the mode at $q_{\vert \vert}=G$ continues to mix two virtual phases with free energies close to the the free energy of
phase 1, i.e. phases 0 and 2. But the mode at $q_{\vert \vert}=2G$ mixes two virtual phases with
 widely different free energies: phase 0 (with free energy close to that of phase 1)
  and phase -1 which has a very low 
 (virtual) ordering temperature $T_{-1}$. Similarly, in phase $N=0$,  both modes 
 at $q_{\vert \vert}=G $  and $2G$ mix a low energy phase
 (resp. $N=1$ and $N=2$) with high energy ones (resp. $N=-1$ and $N=-2$).
  As the  free energy difference between phase 
 $N=0$ and phase $N=1,2$ increases rapidly as the field increases, the MR 
 mode energies at $q_{\vert \vert}=G $ and $2G$
 increase rapidly with field. This is seen in Fig.(\ref{Fig:mr4}).
 
The two MR modes which exist in the low field part of phase $N=1$ have 
different field dependences. The high energy mode, with $q_{\vert \vert}=2G$ 
increases its energy monotonically
 with field, at a rate $\simeq 1.2\delta_1 /T $, 
then merges into the single particle continuum for $H \gtrsim7$ T.
 Then only one MR at $q_{\vert \vert}=G$ survives, at lower energy, 
and with a non monotonic field dependence. In fact the rate of variation of $\omega^{min}_{1,1}/2\delta_1$ is also large,
$\simeq -1.8/T$ for $H\gtrsim 7.5$ T. For $H\lesssim 6.5$ T, $d\omega^{min}_{1,1}/dH\simeq .8\delta_1 /T$.

At the transition from $N=1$ to $N=0$, at $H\simeq 8.25$ T,
 two MR 
modes appear discontinuously. The MR at $q_{\vert \vert}=G$ is
 $21 \%$ 
higher in energy in phase $0$, and that at $q_{\vert \vert}=2G$ is
 close
 to the single particle continuum, at $.95 \times 2\delta_0$. In fact 
this latter mode is also connected to the "contamination" of the Ribault
phase at $N=-2$, since it connects $T_2$ and $T_{-2}$ as described by Eq.(\ref{Tc})\cite{pl97}.
 Both MR energies increase rapidly with $H$, at a rate $\simeq \delta_0/T$ 
for the low energy mode, $1.6\delta_0/T$ for the high energy one.       
 
     \   
 The last surviving MR dissolves into the single particle continuum for $H\gtrsim 12.5$ T.
   For larger $H$ 
  the only collective modes left
 are the usual Goldstone modes at $q_{\vert \vert }\ll G$. Then  the only specific features
  of the UQC left are
  small single particle energy gaps, at most an order of magnitude smaller than the FS gap, 
  at large energies
 $\hbar \omega_c,\;
 2\hbar \omega_c$, etc..
 
              \   
 The low energy physics of the $N=0$ phase for $H\gtrsim 12.5$ T is thus identically that of
  an insulating SDW phase with perfect nesting FS.

The actual field values for which the MR mode drowns in the
 continuum in phase $N=0$ cannot be taken too seriously, since for that 
range of field values, the approximation $2\delta_0/\hbar \omega_c \ll 1$ breaks down completely. However, 
 the behaviour found  above
holds qualitatively since the approximation becomes good again at higher fields.  

In this section, besides  proving that within the RPA, more than one MR mode is usually alive in the single particle gap, 
at $q_{\vert \vert}=G$ or $2G$ (possibly $3G$, or $5,6,7, 8G$ etc. due to the vicinity of
 Ribault phases, as we shall see in the next section) and computing
 their energies as a function of field, we have derived for the first time numerical estimates of the very large
  effective mass anisotropy of the MR. We have described how, as the magnetic field increases, UQC phases at small 
  quantum numbers progressively lose their MR and, above a threshhold field, retrieve a low energy excitation spectrum 
  of a conventional SDW with perfect nesting. In the following,
   we examine how the situation changes in the case of the
  Ribault Phase.

\subsection{MR modes of the Ribault Phase and neighbouring phases}

\begin{figure}[ri3a]
\epsfxsize=8.6cm
\epsfysize=8cm
\epsfbox{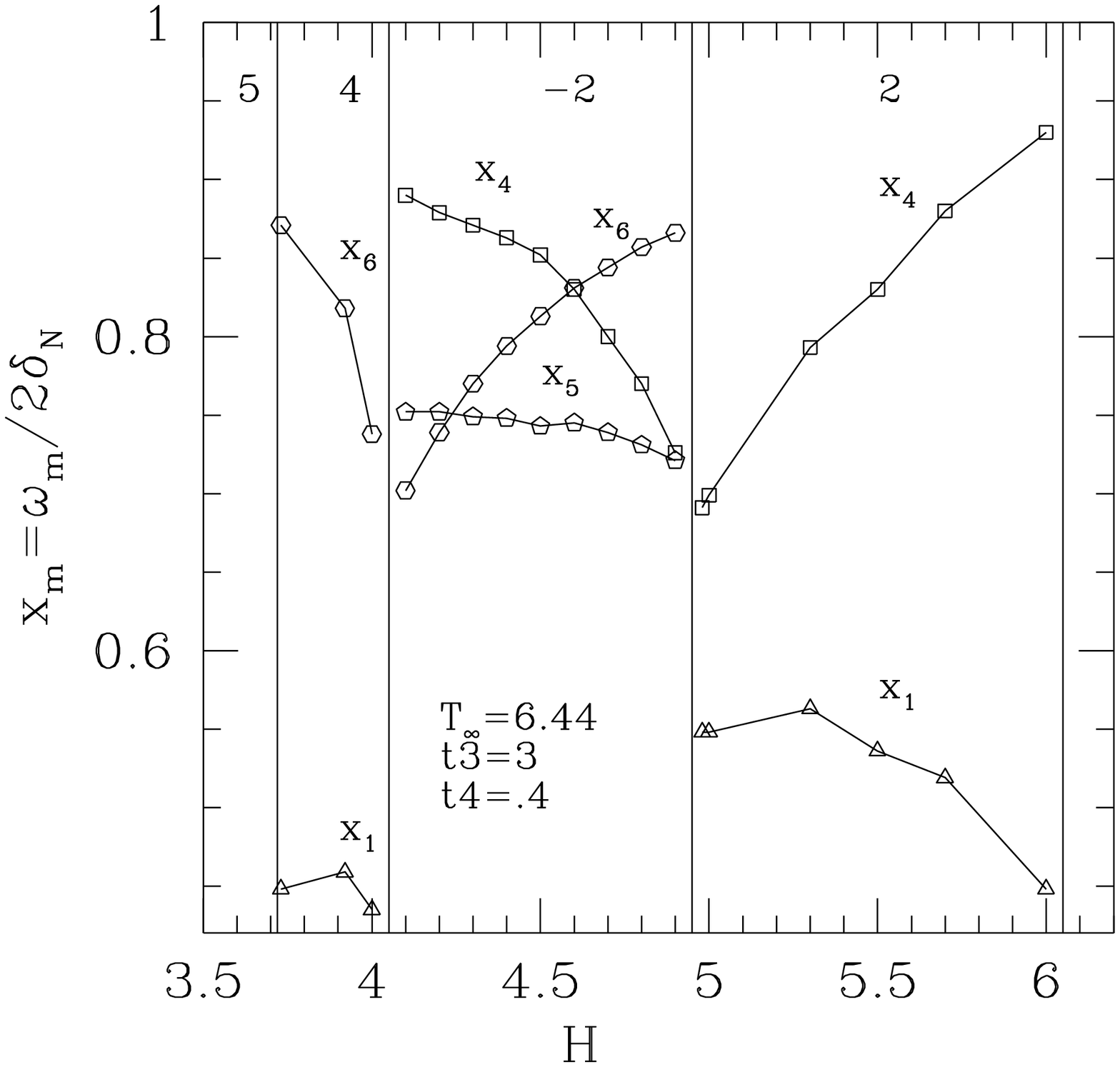}
\caption{Spectrum of MR energy minima in the Ribault phase $N=-2$ for the following 
Quantum Hall number sequence:6,5,4,-2,2,1. Within phase $N=-2$, the MR modes at 4$G$  (6$G$)
mix a low free energy virtual phase  $N=2$ (N=4) with a high free energy one, $N=-6$ (N=-8).
 The almost field independent mode at 5$G$ stems from the coupling to phase $N=3$. The difference
 in free energy between phases $N=3$ and $N=-2$ is almost constant.}
 \label{Fig:ri3a}
             \end{figure}

We now turn to the Quantum Hall sign inversion phenomenon. 
Specifically, we are interested in the following Quantum Hall 
number sequence: ...6,5,4,-2,2,1, which corresponds to the 
UQC cascade shown on Fig.(\ref{Fig:1a})(Notice that no phase $N=3$ 
is thermodynamically stable in Fig.(\ref{Fig:1a})).
 Investigating the MR spectrum of
phase $N=-2$, we now have, from  Eq.(\ref{rot2}), the following possibilities:
\begin{itemize}
\item  Both $N+m$ and $N-m$ are $\leq 0$. Then $T_{N\pm m}$ are so small that 
no MR mode can exist within the single particle gap. This preclude MR modes 
to exist for $m=1$ or $m=2$.
\item  Or $(N+m)(N-m)<0$. Then  one of the virtual transition 
lines, say $T_{-2+m}$ may be close
to $T_{c-2}(H)$, but the other one is very close to zero. Then a MR mode may exist
around $q_{\vert \vert}=mG$, with relative energy 
$\omega_m/2\delta_2\geq 1/\sqrt2$. 
\end{itemize}
As a consequence,
the MR spectrum in the Ribault Phase $N=-2$ exhibits the
 following qualitative features \cite{pl97}: 
 there are no modes at $q_{||}=G$, or $2G$ but two higher energy 
  modes exist at $q_{||}=4G$ and
$q_{||}=6G$; they have large magnetic field dependences of opposite sign; the two 
modes cross roughly in the
 center of phase $-2$ in the phase diagram. In addition, we find a third, almost field 
 independent MR at $q_{||}=5G$
which was not described in ref.(\cite{pl97}). In this latter work, Hall sign inversion 
was studied among a complete
 sequence of integers, as opposed to the sequence where one, or more than one,
  integer is missing, such as the sequence
discussed here: $...6,5,4,-2,2,1$.  In the latter case the mode $q_{||}=5G$  
 is due to the existence of a
 virtual phase at $N=3=-2+5$ the free energy of which is only slightly above that 
 of phase $-2$: the two corresponding 
 lines are practically parallel within the whole $N=-2$ phase, while the 
 "associate" phase at $N=-7=-2-5$ has very 
 large free energy. i. e. very small values of $T_{-7}$. This is another case of 
 contamination from a nearly stable 
 phase. As discussed above, 
   the corresponding MR mode is then field independent, and 
 its value relative to the 
 single particle gap of phase $-2$ is
close to $1/\sqrt 2$. See Fig.(\ref{Fig:ri3a}).
 Another feature apparent in Fig.(\ref{Fig:ri3a})
is the qualitatively different behaviour of modes at $G$ and $4G$ (resp $6G$) as a function of magnetic field
in phase $N=2$ (resp $N=4$). The reason for this difference in behaviour is that the mode at $G$ couples
two virtual phases with low free energies (e.g. phases 1 and 3 for the mode at $G$ in phase 2);
 the latter vary with opposite sign as a function of magnetic field. On the other hand the mode at $4G$ in phase 2
 couples two phases with very different free energies: low energy phase -2 and 
 high energy phase 6. Then according to Eq.(\ref{Tm}), 
 the low free energy virtual phase dictates its field dependence to the MR mode.

\begin{figure}[-4]
   \epsfxsize=8.6cm
   \epsfysize=8cm
   \epsfbox{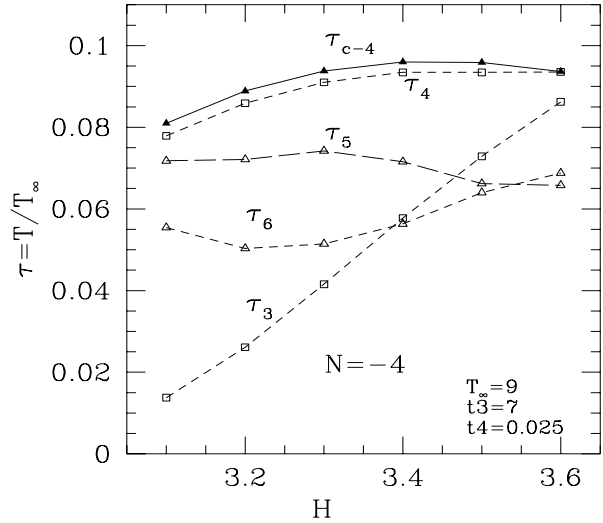}
   \caption{Network of virtual transition lines within phase $N=-4$ for the Quantum Hall number sequence
   6,5,-4,3,2,-2,1. The peculiar  behaviour of MR mode 10 in Fig.(\ref{ri3b}) with field
   stems from the non monotonic variation of $T_6(H)$. MR mode 8 in Fig.(\ref{ri3b}), on the contrary 
   is field independent because the two curves $T_{c,-4}$ and $T_4$ stay close to one another.}
   \label{-4}
               \end{figure}

\begin{figure}[ri3b]
   \epsfxsize=8.4cm
   \epsfysize=8cm
   \epsfbox{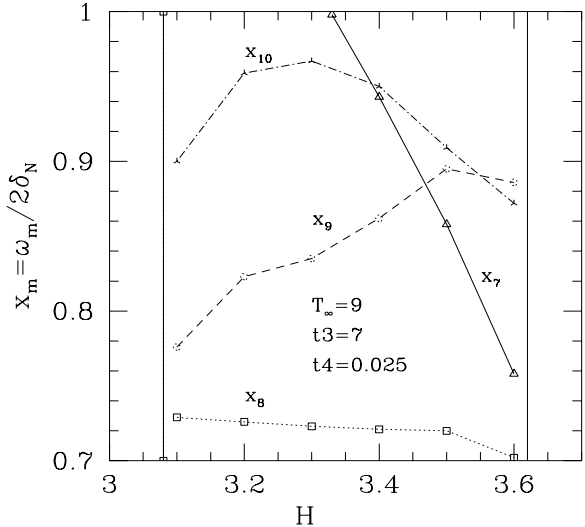}
   \caption{Spectrum of MR energy minima in the Ribault Phase -4 for the Quantum Hall number sequence
   6,5,-4,3,2,-2,1. The loci of MR minima are one order of magnitude further away from the origin than the 
   lowest MR minimum at $q_{\vert \vert}=G$ in the generic positive phase.}
   \label{ri3b}
              \end{figure}

 It is obvious that even richer MR spectra are expected when the FISDW phase diagram
 has more sign reversals. The closer those sign reversed phases in parameter space, the more
 complex the MR spectrum: this is because there are more virtual UQC phases
  with free energies close to that of the stable one. See for example the
   network of virtual transition lines within phase $N=-4$ in Fig.(\ref{-4}).
   This network of virtual transition lines gives rise to the
    MR spectrum shown in Fig.(\ref{ri3b}).
   
   \section{Conclusion}
   
   In view of the renewed confidence in the Quantum Nesting Model as a valid description of the 
   Field Induced Spin Density Wave phenomenon in Bechgaard salts, we have developped the 
   theoretical consequences of the model as regards the collective modes of the UQC.
   
   Our effort has developped in two main directions:
   \begin{itemize}
      \item  We give detailed numerical predictions regarding the collective mode spectrum
 when the phase diagram has no FISDW with negative Quantum Hall number (monotonic Quantum Hall 
 sequence).
 In that case we have proved numerically the existence, within the RPA, of a second MR mode
 with $q_{\vert \vert}=2G$ within the single particle energy gap, in the generic case. Both
 low energy modes at $G$ and $2G$ are almost field independent in the whole domain
  of stability of a given phase. In addition, a strongly field dependent mode at larger parallel
  wave vector may propagate within the single particle gap because of "contamination" from
  a virtual phase with negative Quantum Hall number, when the latter is almost stable. 
  The mass anisotropy of the MR modes are computed for the first time. The parallel to 
  in-plane transverse mass ratio is of order $10^3$.
  
  Then we have studied the MR spectrum of phases with quantum numbers 1 and 0, i.e. phases at large field
  (typically $H\gtrsim 6T$). 
  We show that the two low energy modes become field dependent, their energies increase with field
  and, in the case of the $N=0$ phase, they merge in the single particle excitation continuum.
  Above a threshold field which we have computed, the low energy physics of the FISDW is that
  of a perfectly nested SDW phase.
  \item We have turned to the case of the UQC phase cascade with Quantum Hall sign inversion. This
  unique phenomenon of the Quantum Hall Effect physics,
    observed only in the Bechgaard salts, leads to
  a different structure in the collective mode MR spectra of the sign reversed phases. There
  are no modes with $q_{\vert \vert}=G$ or $2G$; there are modes centered at larger
   wave vectors, with larger energies, and with different
    field dependences; in the case of cascades
   with various sign reversals, there might be four, or more, modes
    in the single particle gap with wave vectors an order of
   magnitude larger than in the monotonic cascade case.
   
   In the course of this latter study, which involves two poorly known
    small parameters $t_3$ and $t_4$ which
   correct the normal metal electronic dispersion relation, we have shown that the cascade 
   phase diagram,
   besides depending sensitively on $t_3$ and $t_4$\cite{zm} depends also sensitively
    on the electron-electron interaction parameter $T_{\infty}$: a change of this parameter
    by a factor 3 may suppress altogether the sign inversion phenomenon. The metal/UQC instability
    line does not scale with $T_{\infty}$.
    
    We have also found that the model contains the (experimentally observed) possibility of
    metallic re-entrance at fixed temperature as a function of field; this may happen also
    when $t_3=t_4=0$.
    \end{itemize}
    
    Our results are obtained within the RPA, in the "very weak" coupling 
    limit ($2\delta_N / \hbar \omega_c \ll 1$). 
In Bechgaard salts, this ratio is of order .4 below 6 T, and increases with 
field. Between 8 T and 20 T, it is 
actually larger than 1, then it decreases and becomes small at very large fields. 
 Thus, significant corrections 
may be expected, in a weak coupling RPA, (i. e. $\lambda \ll 1$) to all estimates
 derived here, in particular 
between 8 T and 20 T. We have found that the lowest MR energy minima of 
the positive UQC phases could be sometimes 
below $\delta_N$. Those values might be significantly lowered 
by self-trapping effects, which are outside 
the scope of linear response theory. However 
the presence of MR with $x_{N,m}< .5$ does not seem to be 
garanteed for all $N$. The effective masses should increase 
due to MR-MR scattering processes,etc.. The importance 
of those effects will be gauged by comparing experimental values to the estimates
 given here. The latter should be a useful guideline  to
determine the MR parameters.

We have provided a number of new quantitative and qualitative predictions on the
spectrum of FISDW collective modes, derived from the QNM model which should help experimentalists 
in the detection of these original fluctuation modes of an original
electron-hole condensate.


\begin{references}
\bibitem{note} It can be a real long lived excited mode in one of the  energy gaps
of the single particle dispersion relation, around $n\hbar \omega_c$, ($n$ integer).
\end{references}
\end{document}